\documentclass[aps,prl,twocolumn,showpacs,superscriptaddress]{revtex4-1}
\usepackage{amsmath}
\usepackage{epsfig}
\usepackage{color}
\usepackage{endnotes}
\usepackage{natbib}
\usepackage[colorlinks,breaklinks]{hyperref}
\usepackage{nicefrac}
\usepackage{bm}
\usepackage{dcolumn}
\usepackage{graphics}
\usepackage{graphicx} 
\usepackage{tikz} 
\usepackage{amsmath}
\usepackage{amssymb}
\usepackage{color}
\usepackage{changes}
\usepackage{multirow}
\usepackage{lineno}
\newcommand{\bra}{\langle}
\newcommand{\ket}{\rangle}

\newcommand{\bs}[1]{\ensuremath{\boldsymbol{#1}}}

\newcommand{\be}{\begin{equation}}
\newcommand{\ee}{\end{equation}}
\newcommand{\bea}{\begin{align}}
\newcommand{\eea}{\end{align}}
\newcommand{\beqa}{\begin{eqnarray}}
\newcommand{\eeqa}{\end{eqnarray}}

\newcommand{\kvec}{\bs{k}}
\newcommand{\pvec}{\bs{p}}
\newcommand{\Pvec}{\bs{P}}

\newcommand{\qvec}{\bs{q}}
\newcommand{\enr}{{\epsilon}}

\begin{document}


\title{Energy and momentum dependence of nuclear short-range correlations - 
Spectral function, exclusive scattering experiments and the contact formalism}

\author{Ronen Weiss}
\affiliation{The Racah Institute of Physics, The Hebrew University, 
             Jerusalem, Israel}
\author{Igor Korover}
\affiliation{Department of Physics, NRCN, P.O.B. 9001, Beer-Sheva 84190, Israel}
\author{Eli Piasetzky}
\affiliation{School of Physics and Astronomy, Tel Aviv University, Tel Aviv 69978, Israel}
\author{Or Hen}
\affiliation{Massachusetts Institute of Technology, Cambridge, Massachusetts 02139, USA}
\author{Nir Barnea}
\email{nir@phys.huji.ac.il}
\affiliation{The Racah Institute of Physics, The Hebrew University, 
             Jerusalem, Israel}

\date{\today}

\begin{abstract} 
Results of electron-induced one- and two-nucleon hard knockout reactions, $A(e,e'p)$ and $A(e,e'pN)$,
in kinematics sensitive to nuclear short-range
correlations, are studied using the nuclear contact formalism. 
A relation between the spectral function and the nuclear contacts is derived and used
to analyze the dependence of the data on the initial 
energy and momentum of the knocked-out proton. 
The ratio between the number of emitted  
proton-proton pairs and proton-neutron pairs is shown to depend predominantly on 
a single ratio of contacts.
This ratio is expected to present a deep minima in the initial 
energy and momentum plane,
associated with the node in the proton-proton wave function.

The formalism is applied to analyze data from recent
$^4$He and $^{12}$C electron-scattering experiments performed at Jefferson laboratory.
Different nucleon-nucleon potentials were used
to asses the model-dependence of the results. 
For the ratio of proton-proton to proton-neutron pairs in $^4$He,
a fair agreement with the experimental data is obtained using the two potentials,
whereas for the ratio of proton-proton pairs to the total knocked-out
protons in $^{12}$C, some of the features of the theory are
not seen in the experimental data. Several possible explanations for this 
disagreement are discussed. It is also observed that the spectral function at specific
domains of the momentum-energy plane is sensitive to the nucleon-nucleon interaction.
Based on this sensitivity, it might be possible to constrain 
the short range part of the nuclear potential using such experimental data.
\end{abstract}


\maketitle
In order to fully describe nuclear systems, it is necessary to understand
the short-range behavior of interacting nucleons, i.e. 
the implications of few nucleons being close to each other inside the nucleus.
These nuclear short-range correlations (SRCs) have been studied intensively in the
last decades. High-energy and large momentum-transfer 
electron and proton-scattering
experiments show that almost all of the nucleons with momentum
larger than the Fermi momentum are part of an SRC pair, which amount to about 
20\% of the nucleons in medium-size and heavy nuclei 
\cite{FraSar93,Egiyan03,Egiyan06,Fomin12,Tang2003,Piasetzky06,Subedi08,Korover14,HenSci14}.
A dominance of neutron-proton pairs was observed among the different possible pairs
\cite{Tang2003,HenSci14,Piasetzky06,Subedi08,Korover14,Baghdasaryan10,Duer2018_np}.
These conclusions are also supported by theoretical works, in which ab-initio
calculations of momentum distributions in nuclei show a universal high-momentum tail,
similar in shape to the deuteron high-momentum tail 
\cite{Schiavilla07,AlvCioMor08,FeldNeff11,AlvCio13,WirSchPie14,Sargsian2005}.
For more details, see recent reviews \cite{Hen_review,Cio15_review}.

Recently, the nuclear contact formalism, a new approach for analyzing nuclear SRCs,
was presented \cite{WeiBazBar15,WeiBazBar15a,WeiBar17,WeiHen17}.
In this theory, new parameters, called the nuclear contacts, describe 
the probability of finding two nucleons close to each other inside the nucleus. 
The values of these contacts depend on the specific nucleus discussed.
Another important ingredients of this theory are the universal two-body functions
that describe the motion of the SRC pairs. 
These functions can be model-dependent, i.e. depend on the
nucleon-nucleon interaction, however they are identical for all nuclei.
This theory was used previously to derive the nuclear contact relations,
which are relations between the nuclear contacts and different nuclear quantities,
such as the one-body and two-body momentum and coordinate space distributions
 \cite{WeiBazBar15a,WeiHen17},
the photo-absorption cross section \cite{WeiBazBar15,WeiBazBar16},
the Coulomb sum rule \cite{WeiPazBar}, and the correlation function \cite{correlation_func}

The purpose of this paper is to study and analyze
electron-scattering experimental data using the contact
theory. We will focus on hard semi-exclusive and exclusive scattering experiments, in which
one or two emitted
nucleons are measured in addition to the scattered electron
\cite{Subedi08,Korover14,HenSci14,Shneor2007,Erez18,Duer18}.
These measurements, in appropriate kinematics,
are one of the main experimental methods for studying nuclear SRCs, and thus
it is important to have a good theoretical description of their results.

In electron-scattering experiments, under the one-photon exchange approximation,
momentum $\bs{q}$ and energy $\omega$ are transferred to the nucleus
by a virtual photon. If $Q^2 \equiv q^2-\omega^2$ is large enough
($\gtrsim 1.5$ GeV$^2$), the photon is predominantly absorbed by a
single nucleon. This nucleon is knocked out from the nucleus and its momentum
$\pvec_1'$ and energy $\enr_1'$ are measured. Neglecting
final-state interaction (FSI), 
the initial momentum and (off-shell) energy $(\pvec_1,\enr_1)$ of the nucleon 
in the nucleus ground state, before it was knocked out, can be reconstructed
\be
  \pvec_1 = \pvec_1'-\qvec, 
  \hspace{3em}
  \enr_1 = \enr_1'-\omega.
\ee
If the initial momentum $p_1$ is larger than the
typical Fermi momentum $p_F\approx 255$ MeV/c$=1.3$ fm$^{-1}$,
then it is most likely that the knocked-out nucleon was part of an SRC pair.
In this case, an emission of a second nucleon is to be expected. This nucleon is the correlated 
partner. Its final
momentum $\pvec_2'$ equals 
its initial-state momentum inside the nucleus $\pvec_2=\pvec_2'$.

This description indicates that the 
semi-exclusive and exclusive cross sections 
should be proportional to the probability
of finding a nucleon with momentum $\pvec_1$ and
energy $\enr_1$ in the initial state,
which is just the definition of the spectral function
$S^N(\pvec_1,\enr_1)$.
Indeed, it was shown in \cite{DeForest83} that 
within the plane-wave impulse approximation (PWIA),
the $(e,e'N)$ cross section is given by
\be \label{cross_section}
\frac{d^4\sigma}{d\Omega_{k'} d\epsilon'_k d\Omega_{p'_1} d\enr'_1}=
           p_1' \enr_1' \sigma_{eN} S^N(\pvec_1,\enr_1)
\ee
where, $k'_{\mu}=(\kvec',\epsilon'_k)$ is the final electron four-momentum,
$N$ denotes a knocked-out neutron or a proton,
and $\sigma_{eN}$ is the off-shell electron-nucleon cross section.

In the case of high-$Q^2$ two nucleon knockout reactions,
previous theoretical \cite{FraStr1988,Sargsian2005,FraStr2008}
and experimental \cite{Piasetzky06} studies have shown
that the measured cross-section can be factorized in a similar manner
to Eq. \eqref{cross_section}, replacing the one-body spectral function
by the two-body decay function $D_A(\pvec_1, \pvec_2, E_R)$. The latter represents
the probability for a hard knockout of a
nucleon with initial momentum $\pvec_1$, followed by an
emition of a recoil nucleon with momentum $\pvec_2$. $E_R$ is the energy of the $A-1$ system, composed of nucleon 2 and the residual $A-2$ nucleus.
We note that integrating the decay function over all recoil nucleon momenta ($\pvec_2$) yields 
the spectral function.


Under few simple assumptions, which will be presented below,  
the asymptotic high-momentum proton spectral function can be written as
\begin{align} \label{spectral_proton}
S^p(p_1,\enr_1) &= C_{pn}^1 S_{pn}^1(p_1,\enr_1) + C_{pn}^0 S_{pn}^0(p_1,\enr_1)
                       \nonumber \\ &+
                             2 C_{pp}^0 S_{pp}^0(p_1,\enr_1).
\end{align}
Here, $C_{ab}^\alpha$ are the nuclear contacts, that measure
the probability to find a proton-proton ($pp$) pair or a proton-neutron ($pn$)
pair close together, with quantum numbers denoted by $\alpha$, while the functions
$S_{ab}^\alpha$ are the contributions of these pairs to the spectral function.
$\alpha=1$ corresponds to the spin-one deuteron quantum numbers, and
$\alpha=0$ corresponds to the spin-zero s-wave quantum numbers. These
are the main two-body channels of nuclear SRC pairs \cite{WeiHen17}.
Based on the experience with the one-body momentum distribution
\cite{WeiHen17}, Eq. \eqref{spectral_proton} is expected to be valid for $p_1 > p_F$.
The probability
to find a proton with energy $\enr_1$ and large momentum $p_1$,
has contribution from both $pp$ and $pn$ pairs.
The equivalent neutron spectral function is obtained by changing between $n$ and $p$.

The derivation of Eq. \eqref{spectral_proton} starts with the 
definition of the spectral function 
\begin{align}\label{spectral_fun_definition}
   S^N(\pvec_1,\enr_1) 
        &= \bar{\sum_i} \sum_{s_1,f}  \delta(\enr_1+E_f^{A-1}-E_i^A)
    \nonumber \\  & \times
      \left|  \bra \Psi_f^{A-1} | a_{\pvec_1,s_1}    | \Psi_i^A  \ket \right|^2
\end{align}
where $\Psi_i^A$ is the ground state wave function,
$E_i^A=(Am-B_i^A)$ is the ground state energy and $B_i^A$ is its binding energy,
$\Psi_f^{A-1}$ is an $(A-1)$-body eigenstate of the nuclear Hamiltonian
with energy $E_f^{A-1}$,
and $ \bar{\sum}_i$ is an average over the
magnetic projections of the ground state. 
$m$ is the nucleon mass and 
$a_{\pvec_1,s_1}$ is the annihilation operator of a nucleon $N$ with momentum
$\pvec_1$ and spin $s_1$.
$S^p$ and $S^n$ are normalized to the total number of protons and neutrons in the nucleus,
correspondingly, i.e., $\int d\enr_1 \frac{d^3p_1}{(2\pi)^3} S^p(\bs{p}_1,\enr_1)=Z$.

For $p_1\longrightarrow\infty$, neglecting three-body or higher correlations,
the ground state wave function is dominated by an SRC pair
with very large relative momentum $\pvec_{12}=(\pvec_1-\pvec_2)/2$
and can be written as 
\be \label{psi_i_12}
  \Psi_i^A  \xrightarrow[p_{12}\rightarrow \infty]{}
	\sum_{\alpha} \tilde{\varphi}_{12}^\alpha(\pvec_{12})
           \tilde{A}_{12}^\alpha(\bs{P}_{12},\{\pvec_k\}_{k\neq 1,2}).
\ee
This is the basic assumption of the contact theory,
and it was validated using ab-initio calculations \cite{WeiHen17,AlvCio16}.
$\tilde\varphi_{ab}^\alpha$ are universal two-body functions, 
while $\tilde{A}_{ab}^\alpha$ describe the motion of the rest of the particles,
and the pair's center of mass (CM) motion, $\Pvec_{12}=\pvec_1+\pvec_2$.
In this picture, once particle $1$ is removed, particle $2$ is left with high
momentum and can be treated as a spectator. Consequently, we may write
\begin{align}\label{psi_f_12}
  \Psi_f^{A,12} &\equiv a_{\pvec_1,s_1}^\dagger \Psi_f^{A-1}
  = 
    {\cal{N}} \hat{\cal{A}}\left\{ \Psi_f^{A-2}
                          | \pvec_1 s_1 ; \pvec_2 s_2 \ket\right\}.
\end{align}
Here,
$\Psi_f^{A-2}$ is an eigenstate of the $(A-2)$-body nuclear Hamiltonian with 
energy $E_f^{A-2}$, $s_i$ is the spin of particle $i$, $\hat{\cal{A}}$ is the anti-symmetrizing operator,
and ${\cal{N}}$ normalization factor.
It follows that
\be \label{A-1_energy}
E_f^{A-1} 
    = \enr_2 + (A-2)m - B_f^{A-2} + \frac{\Pvec_{12}^2}{2m(A-2)}
\ee
where $\enr_2=\sqrt{p_2^2+m^2}$
is the energy of the second correlated nucleon,
$B_f^{A-2}$ is the binding energy of the $(A-2)$-nucleon system,
and the last term is the contribution of the CM motion of the
$(A-2)$-nucleon system.

Substituting Eqs. \eqref{psi_i_12} and \eqref{psi_f_12}  into
Eq. \eqref{spectral_fun_definition}, and  
assuming that the $(A-2)$-nucleon binding energy is narrowly distributed
around a central value $\bar{B}_f^{A-2}$, we arrive at
Eq. \eqref{spectral_proton}. For a pairs of nucleons $ab$, the SRC functions $S_{ab}^\alpha$ are given by
\begin{align} \label{S_ij}
  S_{ab}^\alpha= \frac{1}{4\pi} \int \frac{d\pvec_2}{(2\pi)^3}
          \delta(f(\pvec_2)) 
    \left| {\tilde{\varphi}_{ab}^{\alpha}}(|(\pvec_1-\pvec_2)/2|)\right|^2 
           n_{ab}^{\alpha}(\pvec_1+\pvec_2)
\end{align}
and
\be \label{delta_func}
 f(\pvec_2) = \enr_1+\enr_2-2m+(B_i^A-\bar{B}_f^{A-2})
        +\frac{(\pvec_{1}+\pvec_2)^2}{2m(A-2)},
\ee
where $n_{ab}^\alpha(\Pvec)$, the CM momentum distribution of the
SRC pair, is given by
$ C_{ab}^\alpha n_{ab}^\alpha(\Pvec) = 
   \bra \tilde A_{ab}^\alpha (\Pvec)|\tilde A_{ab}^\alpha (\Pvec)\ket$.
In practice, it can be assumed that all SRC pairs have similar
CM distribution $n_{CM}(\Pvec)$, which 
we shall take as a three-dimensional Gaussian with a width $\sigma_{CM}$
\cite{CioSim96,Erez18,Colle14}.
The spectral functions $S_{ab}^\alpha$ are expected to be almost identical
across the table of nuclides, as the CM and binding energy corrections are
relatively small for nuclei heavier than $^{12}$C \cite{Erez18}.

The delta function in Eq. \eqref{S_ij}
can be used to eliminate the integration over the angles, and 
$ S_{ab}^\alpha$ can be obtained through
numerical integration over $p_2$, without
further approximations. In this integration we also require that
$|(\pvec_1-\pvec_2)/2|>p_F$.
Alternatively, we can continue analytically if we
replace the CM term of Eq. \eqref{delta_func} by its mean value
$\bar{T}_{CM}^{A-2} = \langle P_{12}^2 \rangle / 2m(A-2) = 3 \sigma_{CM}^2 / 2m(A-2)$.
This should be a good approximation for small values of $\sigma_{CM}$ or large values of $A$.
Then, the delta function can be used to fix the magnitude of $\pvec_2$, given by
\be \label{p20_expression}
   p_2^0(\enr_1)=\sqrt{\left[ 2m-\enr_1-(B_i^A-\bar{B}_f^{A-2}) -\bar{T}_{CM}^{A-2} \right]^2-m^2}.
\ee
We can also see that if the CM momentum distribution $n_{CM}(\bs{P})$ has a
zero width, i.e. $n_{CM}$ is a delta function which dictates
$\pvec_2=-\pvec_1$, the spectral function becomes simply a delta function,
centered around
\be \label{zero_CM}
\enr_1 = 2m - \sqrt{p_1^2+m^2} - (B_i^A-\bar{B}_f^{A-2}).
\ee
According to Eq. \eqref{p20_expression}, the momentum magnitude $p_2$
of the second-emitted nucleon in $A(e,e'pN)$ experiments
depends only on the initial energy $\enr_1$ but not on the initial momentum $p_1$
of the knocked-out proton. This might not seem reasonable at first glance,
since we expect that $\pvec_2 \approx -\pvec_1$  \cite{Tang2003,Piasetzky06}.
But, if one substitutes the value of $\enr_1$ of Eq. \eqref{zero_CM}
together with $\bar{T}_{CM}^{A-2}=0$,
into Eq. \eqref{p20_expression}, we obtain $p_2^0=p_1$, as expected.
For a given $p_1$, the value of $\enr_1$ of Eq. \eqref{zero_CM} should be close to a maximum
point in the spectral function, and thus most experimental data is centered around 
such values of $p_1$ and $\enr_1$, leading
to the observation of $\pvec_2 \approx -\pvec_1$.
If sufficient experimental
data of exclusive experiments in other domains
of the momentum-energy plane will be available, it might be possible
to see the energy dependence of $p_2^0$ and compare it to Eq. \eqref{p20_expression}.
We expect for corrections to this relation due finite $A$,
the distribution of the $B_f^{A-2}$ around the mean value $\bar{B}_f^{A-2}$,
and FSI effects.

\begin{figure}\begin{center}
\includegraphics[width=8.6 cm]{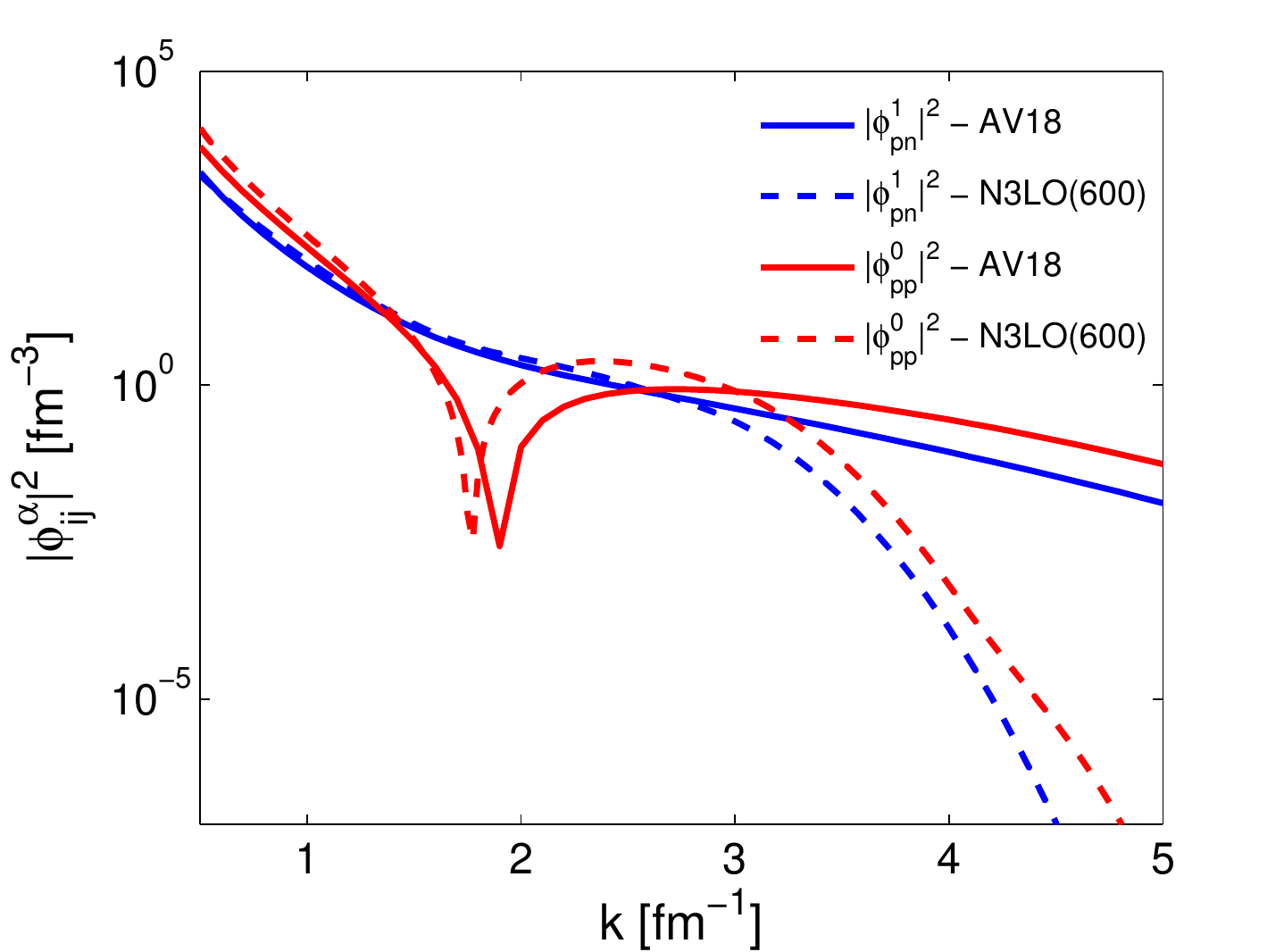}
\caption{\label{two_body_func} 
The universal two-body functions calculated
using two different potentials, for deuteron $pn$ pairs
and s-wave $pp$ pairs.  The functions are normalized such that 
$\int_{p_F}^\infty |\varphi_{ab}^\alpha|^2 d\pvec/ (2\pi)^3 = 1$.
}
\end{center}\end{figure}

To calculate the spectral function we must first calculate the universal
functions $\tilde{\varphi}_{ab}^{\alpha}(\pvec)$. These are
the zero-energy solutions of the two-body Schrodinger equation
for the spin-zero $\alpha=0$ channel, and 
the deuteron wave-function for $\alpha=1$.  
In Fig. \ref{two_body_func} we present the resulting
functions using the AV18 nucleon-nucleon (NN) potential \cite{av18} and 
the chiral EFT NN force N3LO(600) \cite{N3LO}
for the $pp$ spin-zero channel
and the $pn$ deuteron channel. It can be seen that the two potentials
produce similar functions up to the cutoff
value of the N3LO potential  ($p \approx 3$ fm$^{-1}$).
Some differences in the $pp$ functions, like the location of the node,
are observed.

Before presenting our calculations for the spectral function,
we note that $n_{CM}(\bs{P})$ is expected to have a narrow distribution around zero, 
in each axis, with {$\sigma_{CM}\approx p_F/2$}. Therefore, the main contribution to the 
spectral function comes from $\pvec_2$ being anti-parallel to $\pvec_1$.
As can be seen in Fig. \ref{two_body_func}, the $pp$ function has 
a node around $p_{node} \approx 2 \;\text{fm}^{-1}$, and thus we expect $S_{pp}^0$
to have a minimum for
\be \label{node}
\frac{p_1+p_2^0(\enr_1)}{2} = p_{node}.
\ee

\begin{figure} \begin{center}  
    \begin{tikzpicture}
        \node[anchor=south west,inner sep=0] (image) at (0,0) {\includegraphics[width=8.6 cm]
{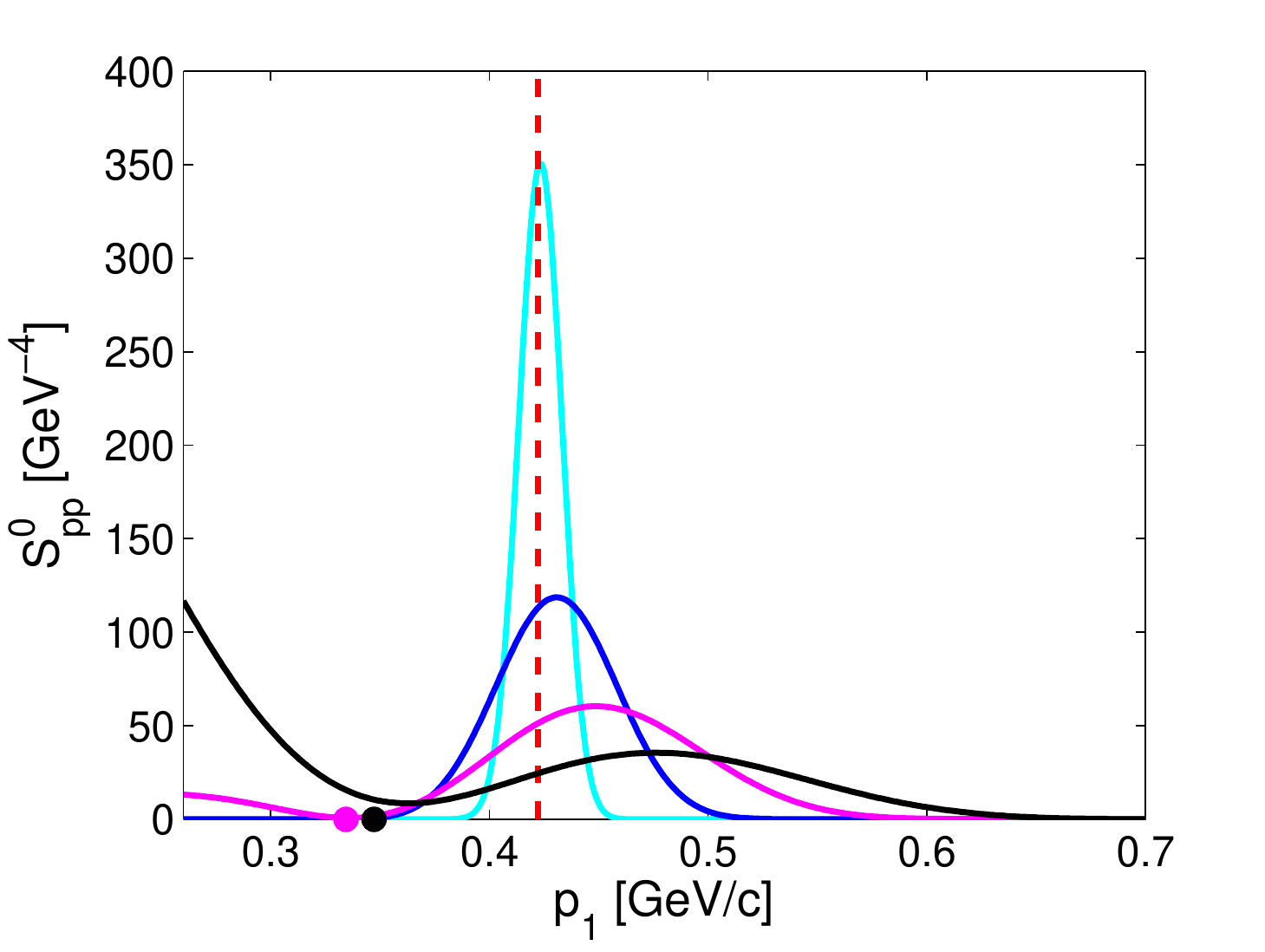}};
        \begin{scope}[x={(image.south east)},y={(image.north west)}]
            \node[anchor=south west,inner sep=0] (image) at (0.45,0.472) {\includegraphics[width=3.81 cm]{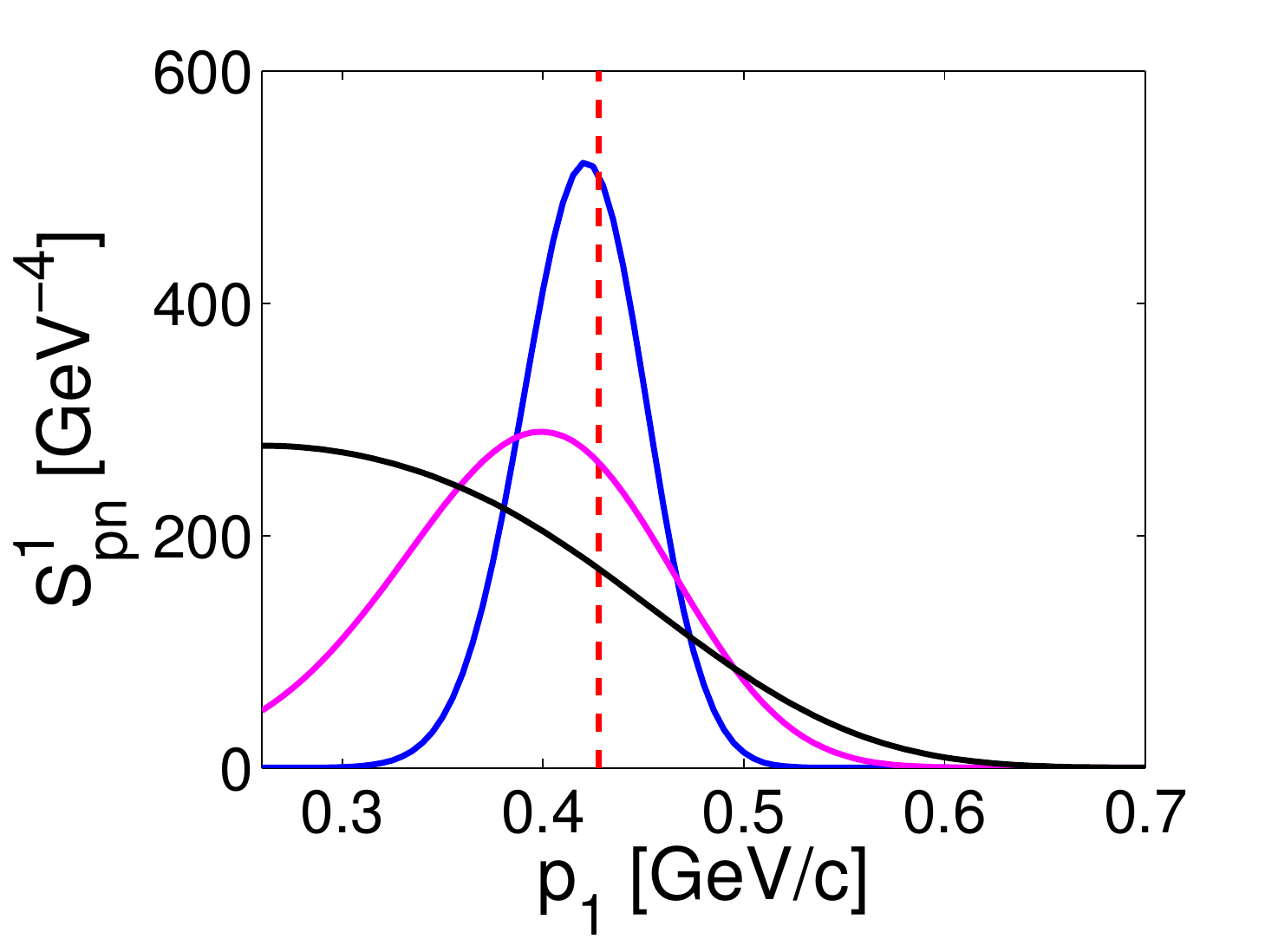}};
        \end{scope}
    \end{tikzpicture}
    \caption{\label{fig_S_ij_momentum} 
$S_{pp}^0$ of $^4$He as a function of $p_1$ for fixed $\enr_1=0.82$ GeV/c,
using the AV18 potential and different values of $\sigma_{CM}$:
10 MeV (cyan), 30 MeV (blue), 60 MeV (magenta) and 100 MeV (black).
The dashed red line is the back-to-back prediction of Eq. \eqref{zero_CM},
and the black and magenta points are the estimated location
of the minimum of $S_{pp}^0$ based on
Eq. \eqref{node}. Inset: the results for $S_{pn}^1$ 
for $\sigma_{CM}=30,\;60$ and $100$ MeV.
}
\end{center} \end{figure}

\begin{figure} \begin{center}  
    \begin{tikzpicture}
        \node[anchor=south west,inner sep=0] (image) at (0,0) {\includegraphics[width=8.6 cm]
{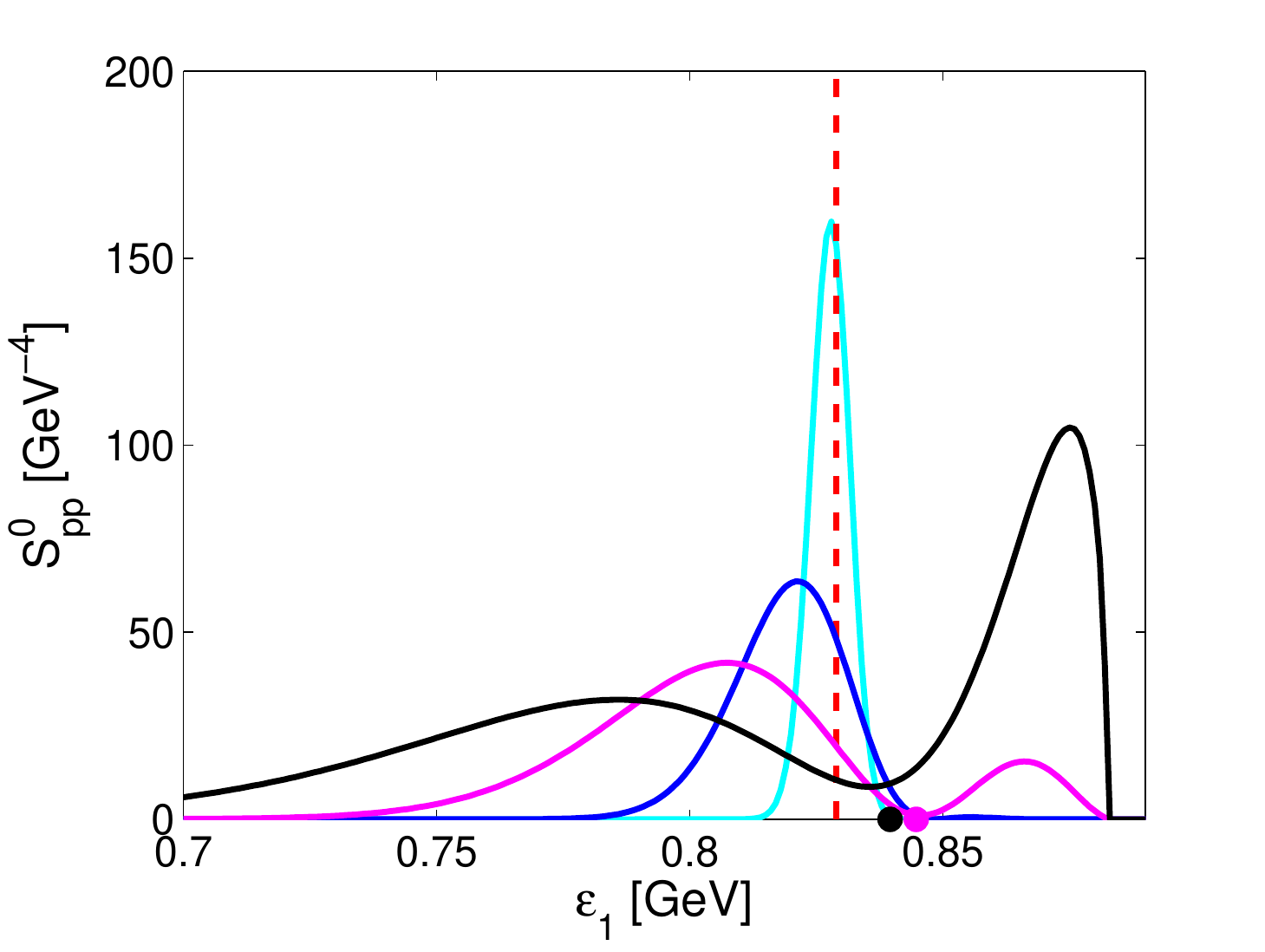}};
        \begin{scope}[x={(image.south east)},y={(image.north west)}]
            \node[anchor=south west,inner sep=0] (image) at (0.148,0.470) {\includegraphics[width=3.81 cm]
{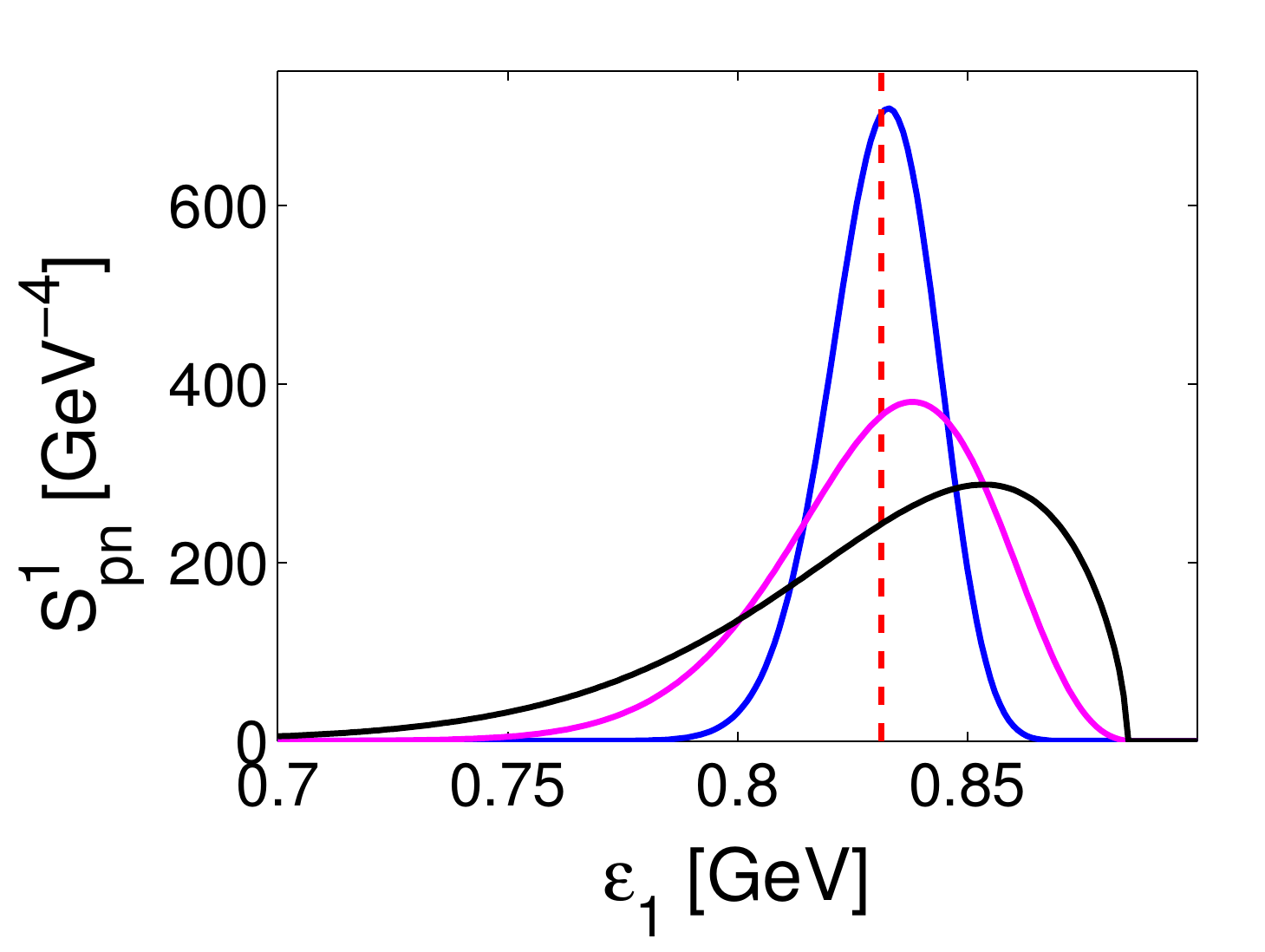}};
        \end{scope}
    \end{tikzpicture}
    \caption{\label{fig_S_ij} 
The same as in Fig. \ref{fig_S_ij_momentum}, but as a function of
$\enr_1$ for fixed $p_1=400$ MeV/c.
}
\end{center} \end{figure}

The calculations of $S_{pn}^1$ and $S_{pp}^0$,
based on Eqs. \eqref{S_ij} and \eqref{delta_func},
are presented in Figs. \ref{fig_S_ij_momentum} and \ref{fig_S_ij},
using the AV18 NN interaction.
In Fig. \ref{fig_S_ij_momentum}, they are presented as a function
of $p_1$, at $\enr_1=0.82$ GeV/c  and different values
of $\sigma_{CM}$. In Fig. \ref{fig_S_ij}, the calculations are 
a function of $\enr_1$ at $p_1 = 400$ MeV/c.
The calculations were done for $^4$He,
taking $B_i^A$ to be its binding energy and 
$\bar{B}_f^{A-2}$ the binding energy of the
deuteron for the $pn$ case and zero for the $pp$ case.
We note that the experimental extraction of $\sigma_{CM}$
of $^4$He is $100 \pm 20$ MeV \cite{Korover14,Erez18}, in a good agreement with
available theoretical estimations \cite{CioSim96,Colle14}.

Calculations for heavier nuclei are similar, with $\enr_1$ shifted
due to the different values of $B_i^A$ and $\bar{B}_f^{A-2}$.
Similar calculations using the N3LO(600) potential are presented in the
supplemental materials.
It can be seen that for small values of $\sigma_{CM}$, the spectral function is
very close to the zero-CM prediction of Eq. \eqref{zero_CM}, corresponding
to back-to-back SRC pairs. As the CM width is increased, $S_{ab}^\alpha$ deviates
from this back-to-back picture.
In addition, we can see that the $pp$ spectral function has an interesting structure
as it develops two maxima for $\sigma_{CM} > 60$ MeV/c.
This structure reflects the node in the $pp$ function, as predicted in
Eq. \eqref{node}.

\begin{figure}\begin{center}
\includegraphics[width=8.6 cm]{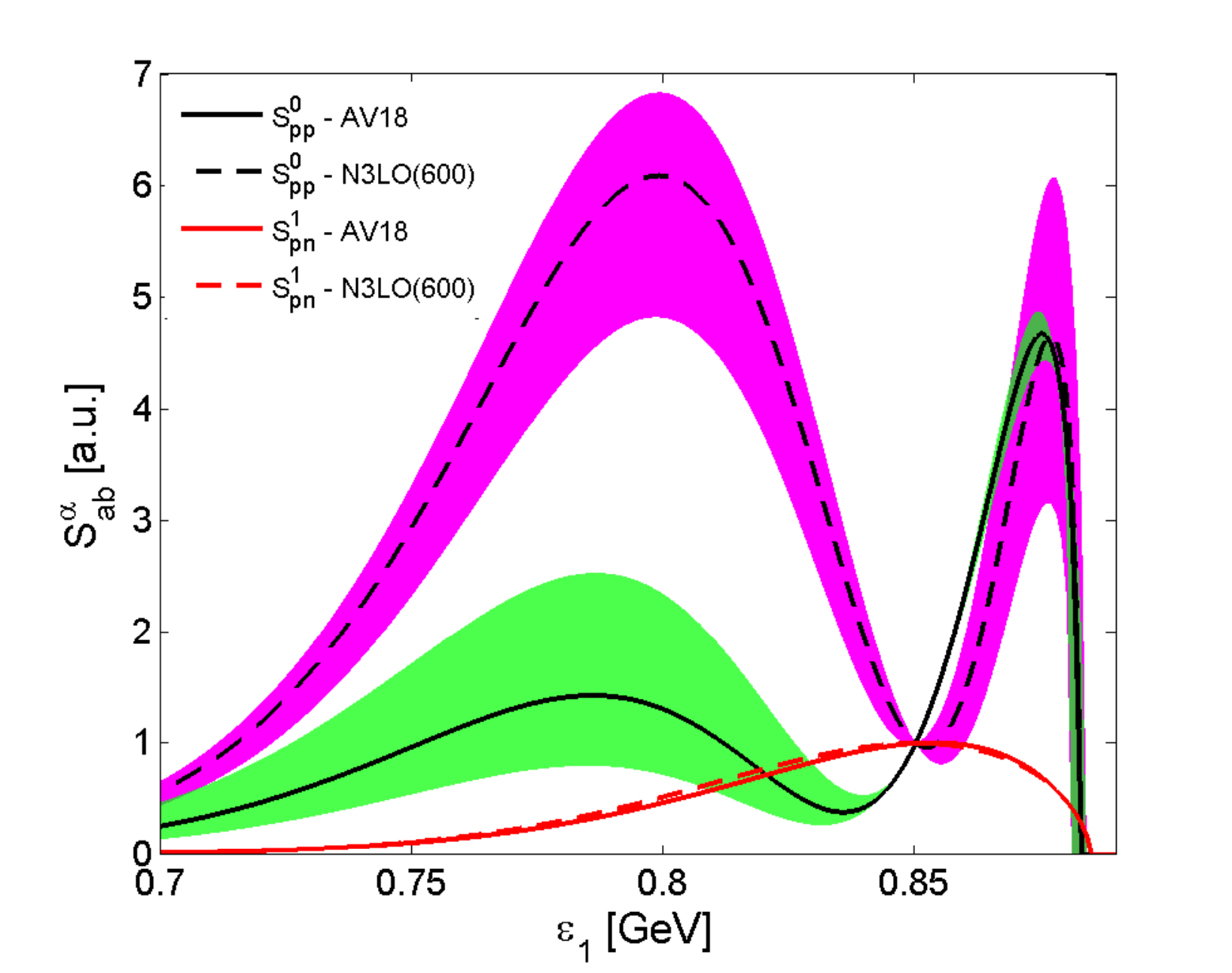}
\caption{\label{fig_S_ij_AV18_N3LO} 
$S_{pp}^0$ and $S_{pn}^1$ for $^4$He as a function of $\enr_1$
for fixed $p_1=400$ MeV/c, normalized to 1 at $\epsilon_1=0.85$ GeV.
The solid and dashed black (red) lines correspond to $S_{pp}^0$ ($S_{pn}^1$) for the AV18 and
N3LO(600) potentials, respectively. The bands around the black lines show the effect of changing the value
of $p_1$ between $390-410$ MeV/c.
The corresponding bands for the red lines are much narrower and are not shown here.}
\end{center}\end{figure}

To compare between the results of the AV18 and N3LO(600) potentials,
we present in Fig. \ref{fig_S_ij_AV18_N3LO} the $^4$He calculations of $S_{pn}^1$
and $S_{pp}^0$, as a function of $\enr_1$ for fixed $p_1=400$ MeV/c and $\sigma_{CM}=100$
MeV/c. Here, the results are normalized to $1$ at $\enr_1=0.85$ GeV.
The bands around the $S_{pp}^0$ results show the effect of changing the value of $p_1$
between $390-410$ MeV.
It is clear that the $S_{pn}^1$ results are very similar for the two potentials,
while the results for $S_{pp}^0$ show significant differences. This is due to the
differences seen in the $pp$ functions presented in Fig. \ref{two_body_func}
around their node.
Based on this sensitivity of $S_{pp}^0$ to the potential, it might be possible
to constrain the short-range part of the $NN$ potential using SRCs experimental data,
as we will further discuss below. We note that $S_{pp}^0$ becomes less sensitive
to the potential for higher or lower values of $p_1$.

It should be noted that our expressions for the spectral function
derived from the contact formalism are similar to the convolution model presented 
by Ciofi degli Atti {\it et al.} in \cite{CioStr91,CioSim96},
and revisited recently in \cite{CioMezMor17a,CioMor17b}. 
The convolution model was shown to agree with ab initio
calculations of the spectral function of $^3$He \cite{CioMezMor17a}. 
Nevertheless, our model differs slightly from the convolution model.
The contact formalism allows us, in principle, to take into account 
contributions from all two-body channels. In this work,
we consider the two leading $np$ two-body
functions, as opposed to a single deuteron function used in the convolution model.
The main contribution comes from the deuteron channel, and we expect 
the additional $np$ channel to have an effect of about $10\%$.
Another small difference is the integration domain of Eq. \eqref{S_ij},
where we included the constrain $|(\pvec_1-\pvec_2)/2|>p_F$, while in the
recent calculations of Ref. \cite{CioMezMor17a} a slightly different constrain was introduced,
leading to a similar effect. Additionally, we use the experimental CM distributions,
as opposed to the ab-initio CM distributions used in the convolution model. Notice also
that we use relativistic expressions for the energy while the convolution model
is completely non-relativistic. A direct comparison between the two models is
presented in the supplemental materials, showing a good agreement for
$^4$He and some differences for $^{12}$C.
 The contact formalism was also shown to agree with ab-initio
calculations of momentum and coordinate-space distributions \cite{WeiHen17}.




Equipped with our contact relation for the spectral function,
we can go back to the exclusive electron-scattering experiments.
One of the main results of these experiments is the ratio between the number
of emitted $pp$ pairs and $pn$ pairs, extracted from the 
$A(e,e'pp)$ and $A(e,e'pn)$ cross sections.
Based on Eq. \eqref{spectral_proton}, we can see that if there is a proton
in some nucleus $A$ with off-shell energy $\enr_1$ and momentum $p_1>k_F$, then it is
part of an SRC pair, which is either a $pp$ pair or a $pn$ pair.
The ratio of the number of such $pp$ to $pn$ pairs is given by
\be \label{pp_pn_ratio_spectral}
  \frac{\#pp}{\#pn}(p_1,\enr_1) 
    =
    \frac{C_{pp}^{0} S_{pp}^0(p_1,\enr_1)}
         {C_{pn}^{1} S_{pn}^1(p_1,\enr_1)  +  C_{pn}^{0} S_{pn}^0(p_1,\enr_1)}.
\ee
For symmetric nuclei ($N=Z$) we expect that 
$C_{pp}^{0} = C_{pn}^{0} \equiv C^0$ \cite{WeiHen17},
and thus this ratio depends only
on a single parameter $C_{pn}^1/C^0$.
We can see that this ratio generally depends on both the
initial momentum of the proton $p_1$ and its energy $\enr_1$.
Within the PWIA,
and based on Eq. \eqref{cross_section}, this ratio can be extracted
from the exclusive-scattering experiments and is given by
$A(e,e'pp)/2A(e,e'pn)$.

The relation of the measured nucleon knockout cross-section ratios to
PWIA calculations and  ground-state energy-momentum densities
relies on the fact that for the high-$Q^2$ kinematics
used in the measurement, according to calculations, reaction mechanisms other than
the hard breakup of SRC pairs are suppressed and any residual effects are significantly
reduced when considering cross-section ratios as oppose to absolute cross-sections
\cite{Hen_review,Arrington2011xs, ColleHen15, Colle2015ena, Boeglin2011mt}. 
The cancellation of reaction
mechanisms in the cross-section ratio steams from the approximate factorization of
the experimental cross-section at high-$Q^2$, which also allows correcting the data
for any remaining effects of FSI and Single-Charge Exchange (SCX)
of the outgoing nucleons using an Eikonal approximation in a Glauber framework
\cite{Cio15_review, Frankfurt1996xx,Colle14,Dutta2013}. 
The experimental data discussed in this work
is already corrected for such effects \cite{Subedi08,Korover14,HenSci14,Shneor2007}.
It should be noted that these corrections were
verified experimentally, see discussion in
\cite{Colle2015ena,Hen:2012yva,Hen_review,Arrington2011xs, Dutta2013,
Frankfurt2001,Pieper1992}.

The $\#pp/\#pn$ ratio was extracted from exclusive-scattering
experimental data for
$^4$He \cite{Korover14} and $^{12}$C \cite{Subedi08,Shneor2007}.
In these experiments, the main focus was the dependence of these ratios
on the initial momentum $p_1$, and not the dependence on $\enr_1$.
In both experiments, the ratios were measured in several
kinematical settings, each corresponding to specific central values of
$p_1$ and $\enr_1$. The momentum-dependence of the ratio was
highlighted, but the effects of the initial energy $\enr_1$ were not
discussed.
This discussion is also missing in previous theoretical works 
that used the momentum distribution as a starting point to predict
the $\#pp/\#pn$ ratio
\cite{AlvCio16,WeiHen17,Ryc15,NeffFeldHor15}. 
The study of this ratio, and  
SRC pairs in general, should be extended to include the full
energy and momentum $(\enr_1,p_1)$ dependence. 

Using Eq. \eqref{pp_pn_ratio_spectral} we can predict the value
of the $\#pp/\#pn$ ratio as a function of both $p_1$ and $\enr_1$,
for any nucleus, if the values of the contacts and $\sigma_{CM}$ for this nucleus are
known. The values of the contacts for several nuclei with mass number up to $A \leq 40$ were
extracted recently \cite{WeiHen17} using variational Monte Carlo (VMC) two-body densities
in momentum and coordinate space \cite{WirSchPie14,Wiringa_CVMC}, 
calculated using the AV18 NN potential and the 
Urbana X (UX) three-nucleon force \cite{ubx}.
We will focus here on $^4$He and $^{12}$C, for which the experimental
data is also available. As mentioned before,
for symmetric nuclei as these, the $\#pp/\#pn$ ratio depends only on one contact ratio.
We use the available experimental data of Refs. \cite{Korover14}
to fit this ratio of contacts for $^4$He, utilizing Eq. \eqref{pp_pn_ratio_spectral}.
For $^{12}$C, we fit the ratio of contacts to the $\#pp/\#p$ ratio
of Ref. \cite{Shneor2007}, which will be discussed below.
The fitted values for $^4$He and $^{12}$C 
are given in Table \ref{fitted_contacts}, using the AV18,
the N3LO(600) 
and the local chiral N2LO \cite{GezSch13,GezSch14} potentials,
for the calculation of the spectral function,
the experimental estimate $\sigma_{CM}(^4\rm{He})=100\;\rm{MeV}$ \cite{Korover14},
and $\sigma_{CM}(^{12}\rm{C})=143\;\rm{MeV}$ \cite{Tang2003,Shneor2007,Erez18},
and the relevant bound-state energies for $B_i^A$ and $\bar{B}_f^{A-2}$.
The local N2LO chiral potential includes two cutoffs,
$R=1.0$ fm and $R=1.2$ fm, denoted here by N2LO(1.0) and N2LO(1.2),
respectively.
Previously extracted contact values,
using the AV18 NN potential and the UX three-body force, are also given in the table,
and agree with the AV18 ratio extracted here.
This ratio of contacts $C_{pn}^1/C^0$ gives us the ratio between the total
number of SRC $pn$ pairs in the deuteron channel and the number of SRC $pp$ pairs.
Only the values in the first column of table  \ref{fitted_contacts}
are used in the reminder of this paper.

The extracted contact ratio using N3LO(600), also shown in table
\ref{fitted_contacts}, is larger than the one obtained using AV18,
which shows that this ratio is model dependent.
The main source for this model dependence is the sharp fall of the
N3LO(600) $|\tilde{\varphi}_{ab}^\alpha|^2$ functions for $p>3$ fm$^{-1}$
(Fig. \ref{two_body_func}). This reduces significantly the 
number of SRC $pp$ pairs, i.e. the value of $C_{pp}^0$,
 because the contribution of  $p>3$ fm$^{-1}$ is small, while 
the AV18 $pp$ function has significant
contribution to SRC pairs for $p>3$ fm$^{-1}$.
We can look on the total number of $pn$ deuteron pairs over $pp$ pairs
with relative momentum restricted to $p_F < p < p_{max}\equiv 3$ fm$^{-1}$, given
by
\be
\frac{C_{pn}^1 \int_{p_F}^{p_{max}} d\pvec |\tilde{\varphi}_{pn}^1(\pvec)|^2}
     {C_{pp}^0 \int_{p_F}^{p_{max}} d\pvec |\tilde{\varphi}_{pp}^0(\pvec)|^2}.
\ee
For AV18 we get a ratio of $32 \pm 8$ for $^4$He, which is much larger
than the ratio of all $p>p_F$ pairs of table \ref{fitted_contacts}.
For N3LO(600) we get
a ratio of $35 \pm 9$ for $^4$He, similar to the original ratio shown in the table.
We can see that the two potentials give consistent values when restricting the 
momentum range to  $p_F < p < 3$ fm$^{-1}$, and the model dependence disappears.
Similar result is obtained also for $^{12}$C.
In this discussion, it is important to distinguish between two $\#pp/\#pn$ SRC
ratios. One is measured in exclusive scattering, given by Eq. \eqref{pp_pn_ratio_spectral},
and depends on both the initial momentum $p_1$ and the initial energy $\enr_1$
of the knocked out proton. The second,
describes the number of $pp$ and $pn$ (deuteron) pairs with relative momentum $p$,
and is given by
$C_{pp}^0 |\tilde{\varphi}_{pp}^0(p)|^2/C_{pn}^1 |\tilde{\varphi}_{pn}^1(p)|^2$.

Regarding the local chiral interactions, for $^4$He, the 
'‘hardest’' chiral interaction, N2LO(1.0), results in a contact ratio
that is very similar to that of the phenomenological AV18 interaction.
Increasing its cutoff to 1.2 fm slightly reduces the contact ratios.
As mentioned above, the softer non-local N3LO(600) interaction
produces a larger contact ratio.
For $^{12}$C, the cutoff dependence of the N2LO interaction is somewhat
less pronounce and they both agree, within uncertainties,
with the AV18 extraction.
As discussed before, some of these differences can be attributed to the
differences in the universal functions, which depend on the potential.
Model-independence is expected for contact ratios of two nuclei, for
the same interaction and two-body channel, as observed in Refs. 
\cite{charge_density,Chen17_EMC}, but not for
the ratios presented in Table \ref{fitted_contacts}. 
Decisive conclusion regarding such model independence is not 
possible here, due to the relatively large uncertainties in the
extracted contact values.

\begin{table}
\begin{tabular}{c| c  c  c  c }
  \hline
\hline                       
A & potential  & (e,e'pN) & k-VMC & r-VMC  \\
\hline
\multirow{4}{*}{$^4$He} &
  AV18 & $20 \pm 5$ & $18.4 \pm 0.8$ & $20.5 \pm 0.2$\\
 & N3LO(600) & $33 \pm 8$  & - & - \\
 & N2LO(1.0) & $19 \pm 5$  & - & -\\
 & N2LO(1.2) & $15 \pm 4$  & - & -  \\
\hline
\multirow{4}{*}{$^{12}$C} &
  AV18 & $ 14 \pm 3$ & $ 12.5 \pm 2 $ & $18.0 \pm 0.2 $\\
 & N3LO(600) & $ 25 \pm 5 $ & - & - \\
&   N2LO(1.0) & $ 19 \pm 4$  & - & - \\
 & N2LO(1.2) & $ 20 \pm 5$   & - & - \\
\hline  
\hline    
\end{tabular}
 \caption{\label{fitted_contacts} The fitted values of the contact ratio
$C_{pn}^1/C^0$ for $^4$He and $^{12}$C. The rows correspond to different
potentials and the columns correspond to different fits. $(e,e'pN)$ is the
fit to the experimental $\#pp/\#pn$ ratio of Ref. \cite{Korover14} for $^4$He,
and to $\#pp/\#p$ of Ref. \cite{Shneor2007} for $^{12}$C, presented in this work.
The k-VMC and r-VMC are fits to VMC two-body densities in momentum and coordinate space,
respectively, taken from Ref. \cite{WeiHen17}.
Only the values in the $(e,e'pN)$ column are used in this paper.
%
%
%
%
}
\end{table}

\begin{figure}\begin{center}
\includegraphics[width=8.6 cm]
{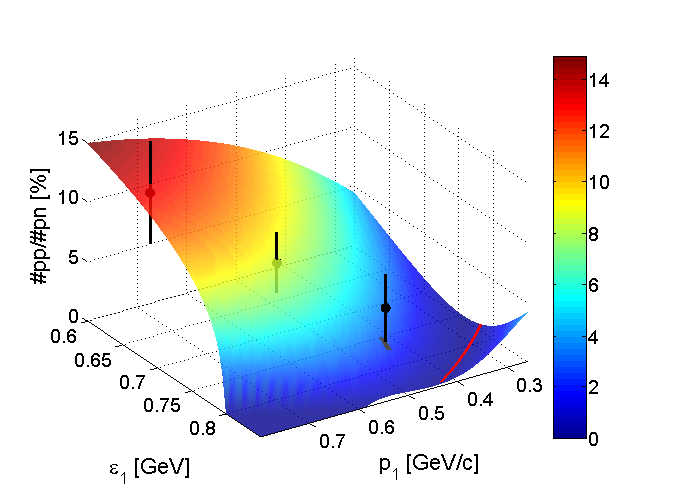}
\includegraphics[width=8.6 cm]
{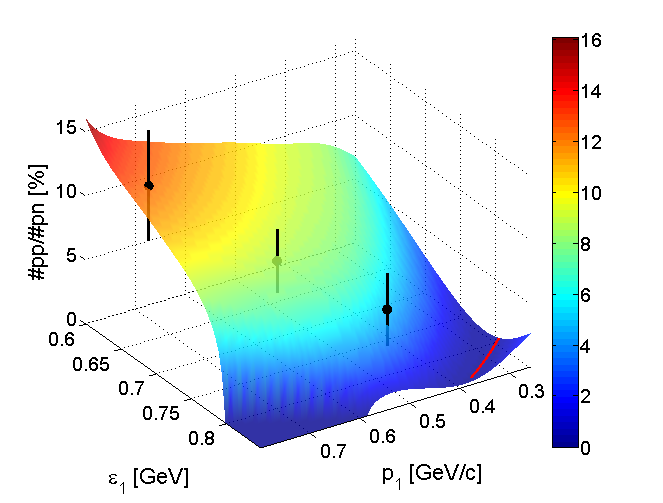}
\caption{\label{pp_pn_surface} 
(Top) The $^4$He $\#pp/\#pn$ ratio as a function of both $p_1$ and $\epsilon_1$,
according to Eq. \ref{pp_pn_ratio_spectral} and the contact ratio fitted 
in this work (table \ref{fitted_contacts}), using the AV18 potential.
The red line is the analytic prediction for a minimal ratio value,
and the black points are the experimental data of Ref. \cite{Korover14}.
The location of experimental points that do not intersect the surface are
indicated by a gray patch on the surface.
(Bottom) The same but using the N3LO(600) potential.
The values of the experimental data in the momentum and 
energy axes are $p_1=0.49 \pm 0.1,\; 0.62\pm 0.09,\; 0.75 \pm 0.08$ GeV/c
and $\enr_1=0.81^{+0.09}_{-0.21},\; 0.74^{+0.11}_{-0.19},\; 0.66^{+0.09}_{-0.21}$
GeV, respectively.
}
\end{center}\end{figure}

Using the fitted contact ratio for $^4$He,
we can now predict the full dependence of the $\#pp/\#pn$ ratio.
The results are presented in Fig. \ref{pp_pn_surface} using the AV18 and N3LO(600) potential.
We can see that the surface describes well the exclusive-scattering experimental
data of Ref. \cite{Korover14} (the black points) using both potentials.
We also include our analytic prediction for the $(p_1,\enr_1)$ points
for which the $\#pp/\#pn$ ratio is minimal (red line), based on Eq. \eqref{node}.
There is a good agreement with the full numerical calculations.
One can see that the available experimental data sits on a diagonal
line in the $(p_1,\enr_1)$ plane, while there is no experimental data 
for substantial parts of this plane. Thus, additional experimental
data, covering the $(p_1,\enr_1)$ plane, is needed to fully investigate the theoretical 
predictions
presented in Fig. \ref{pp_pn_surface}.

Based on Fig. \ref{pp_pn_surface}, it seems that AV18 and N3LO(600) 
predict a similar structure for $\#pp/\#pn$.
This takes us back to Fig. \ref{fig_S_ij_AV18_N3LO},
which showed that $S_{pp}^0$ is sensitive to the NN potential around $p_1 =400$ MeV.
Thus, if the number of SRC $pp$ pairs
will be measured in future exclusive experiments
as a function of $\enr_1$ with fixed $p_1=400$ MeV,
it might be possible to use it to constrain the NN potential.
Since we are discussing $pp$ pairs with high relative momentum, it should be sensitive to
the short distance part of the potential.
Based on the bands presented in Fig. \ref{pp_pn_surface},
we note that the experimental uncertainty of the value of $p_1$ should
not be larger than $10$ MeV, in order to differentiate between AV18 and N3LO(600).

One can also consider the $\#pp/\#p$ ratio, i.e. the number of
correlated $pp$ pairs consisting of a proton with off-shell momentum-energy
$(p_1,\enr_1)$, divided by the total number of such protons.
For $p_1>k_F$, this ratio should be given by
\be \label{pp_p_ratio_spectral}
  \frac{\#pp}{\#p}(p_1,\enr_1) 
    =
    \frac{C_{pp}^{0} S_{pp}^0}
         {2C_{pp}^0 S_{pp}^0 +
        C_{pn}^{1} S_{pn}^1  +  C_{pn}^{0} S_{pn}^0}.
\ee
This ratio was extracted from exclusive scattering experiments
for $^4$He \cite{Korover14} and $^{12}$C \cite{Shneor2007}.
We note that similar corrections to those discussed above (for FSI and SCX)
were already applied to the cross sections to obtain the experimental $\#pp/\#p$ ratio.
These corrections are much more significant here, comparing to the $\#pp/\#pn$ corrections,
and include transparency effects and significant
model-dependent acceptance corrections (of the order of a factor of $10$ for the experimental
data analyzed here).

\begin{figure}\begin{center}
\includegraphics[width=8.6 cm]
{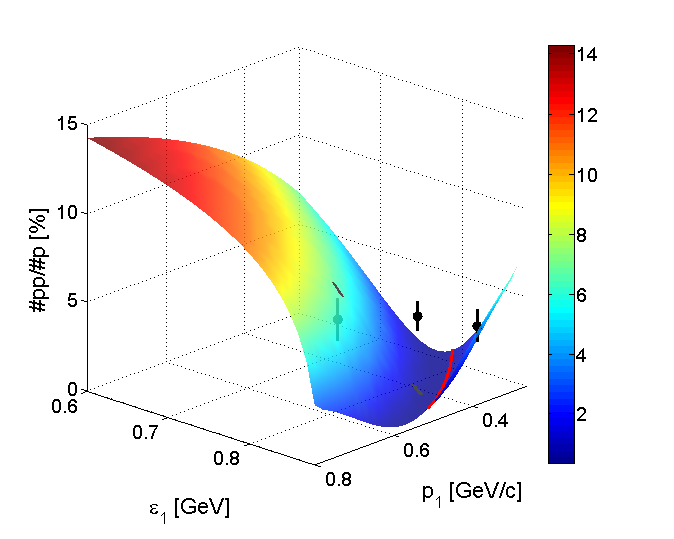}
\caption{\label{pp_p_surface} 
The same as in Fig. \ref{pp_pn_surface}, but for the $^{12}$C $\#pp/\#p$ ratio,
according to Eq. \eqref{pp_p_ratio_spectral}, using the AV18 potential.
The black points are the experimental data of Ref. \cite{Shneor2007}.
The values of the experimental data in the momentum and 
energy axes are $p_1=0.35 \pm 0.05,\; 0.45\pm 0.05,\; 0.55 \pm 0.05$ GeV/c
and $\enr_1=0.86^{+0.04}_{-0.11},\; 0.84^{+0.05}_{-0.15},\; 0.79^{+0.11}_{-0.14}$
GeV, respectively.
}
\end{center}\end{figure}

Fig. \ref{pp_p_surface} depicts the $\#pp/\#p$ ratio for $^{12}$C using the
AV18 potential, based on Eq. \eqref{pp_p_ratio_spectral}
and the contact ratio fitted in this work (table \ref{fitted_contacts}),
compared to the experimental data of Ref. \cite{Shneor2007}.
Here, one can see that while the theory 
predicts a deep minima in the ratio, the experimental data
seems to show a constant ratio of about 5\%.
Similar figure is presented in the supplemental materials
using the N3LO(600) potential.
There are few possible explanations for this disagreement
between our theory and the data. As mentioned above,
the corrections applied to the data in order to obtain the $\#pp/\#p$ ratio are
quite significant. The disagreement shown in Fig. \ref{pp_p_surface}
might indicate that these corrections should be re-examined.
Experimental data which requires smaller corrections can
be useful here, for example using large-acceptance detectors (see e.g. Ref. \cite{HenSci14}). 
It is also possible that the limited statistics and the large bins
of the data presented in Fig. \ref{pp_p_surface}
smears the finer details of the $\#pp/\#p$ ratio, yielding
approximately a constant ratio. If this is the case, to verify the theoretical predictions
of this work, better data is needed.
Finally, corrections to the theory should also be studied, such as the effects of
the energy distribution of the $A-2$ system ($B_f^{A-2}$) around its mean value.

In the supplemental materials, we present the $\#pp/\#p$ ratio also
for $^4$He and the $\#pp/\#pn$ ratio for $^{12}C$, using the same values
of the contacts (table \ref{fitted_contacts}). 
Similar to $^{12}$C, the experimental data for the 
$\#pp/\#p$ ratio of $^4$He \cite{Korover14} seems to indicate
 a constant value for the ratio, while the 
theory shows a different picture.
The single experimental point for the
$^{12}$C $\#pp/\#pn$ ratio is in agreement with the theoretical predictions.
The analysis of the $\#pn/\#p$ ratio is also presented in the supplemental materials
for $^4$He and $^{12}$C. The experimental data for this ratio \cite{Korover14,Subedi08}
includes quite large errorbars and better data is needed to
investigate the theoretical predictions.
Similar analysis using the local chiral N2LO potential
is also presented in the supplementary.

To summarize, the nuclear contact formalism
was used to derive a relation between the nuclear contacts,
describing the probability to find SRC pairs in the nucleus,
and the spectral function. This relation was utilized to analyze
the $\#pp/\#pn$, $\#pp/\#p$ and $\#pn/\#p$ ratios for $^4$He and $^{12}$C,
emphasizing the full dependence in the
$(p_1,\enr_1)$ plane and revealing a richer structure than was assumed so far,
using few different nuclear potentials.
For $\#pp/\#pn$ there is a good agreement with the available experimental data,
extracted from exclusive electron-scattering experiments,
while for $\#pp/\#p$ there seems to be a disagreement. Possible explanations
for this disagreement were discussed. Better experimental data is needed
for $\#pn/\#p$ in order to compare with the theoretical predictions.
The contact ratio $C_{pn}^1/C^0$ for $^4$He and $^{12}$C extracted
using the AV18 potential agrees with previous values, extracted using the same potential.
The contact values seem to depend on the NN interaction, but this
model dependence is resolved if one is looking on a limited high-momentum range.
It was also shown that the contribution of SRC $pp$ pairs to the spectral function
is sensitive to the NN potential, which can be used to constrain the short-range part
of the potential, if appropriate experimental data is available.

A main conclusion of this work is that the full
energy and momentum dependence of exclusive
electron-scattering experiments should be studied, 
experimentally and theoretically, 
in order to obtain a full picture regarding nuclear SRCs.
Further experimental data for the $\#pp/\#pn$, $\#pp/\#p$ and $\#pn/\#p$ ratios and other observables,
for different nuclei, covering the energy-momentum plane, is required for investigating
the predictions presented in this work.

\begin{acknowledgments}
This work was supported by the Pazy foundation,
the Israel Science Foundation and by
the Office of Nuclear Physics of the U.S. Department of Energy
under grant Contract Numbers DE-FG02-94ER40818.
\end{acknowledgments}


\begin{thebibliography}{99}

\bibitem{FraSar93}
L.L. Frankfurt, M.I. Strikman, D.B. Day, M. Sargsyan 
Phys.Rev. C {\bf 48}, 2451 (1993)
\bibitem{Egiyan03} 
K. Egiyan, {\it et al.}, Phys. Rev. C {\bf 68}, 014313  (2003).
\bibitem{Egiyan06} 
K. Egiyan, {\it et al.}, Phys. Rev. Lett. {\bf 96}, 082501 (2006).
\bibitem{Fomin12}
  N. Fomin {\it et al.}, 
  Phys. Rev. Lett. {\bf 108}, 092502 (2012).

\bibitem{Tang2003} 
 A. Tang, {\it et al.}, Phys. Rev. Lett. {\bf 90}, 042301 (2003).
\bibitem{Piasetzky06}
  E. Piasetzky, M. Sargsian, L. Frankfurt, M. Strikman, J.W. Watson, 
  Phys. Rev. Lett. {\bf 97}, 162504 (2006).
\bibitem{Subedi08}
  R. Subedi {\it et al.}, Science {\bf 320}, 1476 (2008). 
\bibitem{Korover14}
I. Korover, {\it et al.}, Phys.Rev.Lett. 
{\bf 113}, 022501 (2014).

\bibitem{HenSci14}%
  O. Hen {\it et al.} (CLAS Collaboration), 
  Science {\bf 346}, 614 (2014).

\bibitem{Baghdasaryan10} 
H. Baghdasaryan {\it et al.},
Phys. Rev. Lett. {\bf 105}, 222501 (2010).

\bibitem{Duer2018_np}
M. Duer {\it et al.} (CLAS Collaboration),
arXiv: 1810.05343 (2018)

\bibitem{Schiavilla07} 
R. Schiavilla, R. B. Wiringa, Steven C. Pieper, and J. Carlson,
Phy. Rev. Lett. {\bf 98}, 132501 (2007).
\bibitem{AlvCioMor08}%
  M. Alvioli, C. Ciofi degli Atti and H. Morita, 
  Phys. Rev. Lett. {\bf 100}, 162503 (2008).
\bibitem{FeldNeff11}
H. Feldmeier, W. Horiuchi, T. Neff, and Y. Suzuki, Phys.
Rev. C {\bf 84}, 054003 (2011)
\bibitem{AlvCio13} 
M. Alvioli, C. Ciofi degli Atti, L. P. Kaptari, C. B. Mezzetti, and H. Morita,
Phys. Rev. C {\bf 87}, 034603 (2013)
\bibitem{WirSchPie14}%
  R. B. Wiringa, R. Schiavilla, S. C. Pieper, J. Carlson, 
  Phys. Rev. C {\bf 89}, 024305 (2014).
\bibitem{Sargsian2005}
M. Sargsian, T.V. Abrahamyan, M.I. Strikman, and L.L. Frankfurt,
Phys. Rev. {\bf C} 71, 044615 (2005)

\bibitem{Hen_review}%
O. Hen, G.A. Miller, E. Piasetzky, and L. B. Weinstein,
Rev. Mod. Phys. {\bf 89}, 045002 (2017)
\bibitem{Cio15_review}%
  C. Ciofi degli Atti, Phys. Rep. {\bf 590}, 1 (2015).

\bibitem{WeiBazBar15} 
  R. Weiss, B. Bazak, and N. Barnea, 
  Phys. Rev. Lett. {\bf 114}, 012501 (2015).
\bibitem{WeiBazBar15a} 
  R. Weiss, B. Bazak, and N. Barnea, 
  Phys. Rev. C {\bf 92}, 054311 (2015).
\bibitem{WeiBar17} 
  R. Weiss and N. Barnea, Phys. Rev. C {\bf 96}, 041303(R) (2017).
\bibitem{WeiHen17}  
  R. Weiss, R. Cruz-Torres, N. Barnea, E. Piasetzky, and O. Hen,
Phys. Lett. B {\bf 780}, 211 (2018)
\bibitem{WeiBazBar16} 
  R. Weiss, B. Bazak, and N. Barnea, 
  Eur. Phys. J. A {\bf 52}, 92 (2016)
\bibitem{WeiPazBar}  
R. Weiss, E. Pazy, and N. Barnea, 
Few-Body Syst \bs{58}, 9 (2017) 
\bibitem{correlation_func} 
R. Cruz-Torres, {\it et. al.}, 
Phys. Lett. B {\bf 785}, 304 (2018)

\bibitem{Shneor2007}
R. Shneor, {\it et al.}, Phys. Rev. Lett. {\bf 99}, 072501 (2007) 
\bibitem{Erez18}   
 E. O. Cohen {\it et al.} (CLAS Collaboration),
Phys. Rev. Lett. {\bf 121}, 092501 (2018)
\bibitem{Duer18}
M. duer {\it et al.} (CLAS Collaboration), 
Nature {\bf 560}, 617 (2018)


\bibitem{DeForest83} 
T. De Forest, Nucl. Phys. A {\bf 392}, 232 (1983)

\bibitem{FraStr1988} 
L. Frankfurt, and M. Strikman,
Phys. Rep. {\bf 160}, 235 (1988).

\bibitem{FraStr2008} 
 L. Frankfurt, M. M. Sargsian, and M. Strikman,
Int. J. Mod. Phys. A {\bf 23}, 2991 (2008)





\bibitem{AlvCio16} 
  M. Alvioli,  C. Ciofi degli Atti, and H. Morita,
Phys. Rev. C {\bf 94}, 044309 (2016)

\bibitem{CioSim96}
  C. Ciofi degli Atti and S. Simula,
  Phys. Rev. C {\bf53}, 1689 (1996). 

\bibitem{Colle14}
C. Colle, W. Cosyn, J. Ryckebusch, and M. Vanhalst,
  Phys. Rev. C {\bf 89}, 024603 (2014).



\bibitem{av18} 
  R. B. Wiringa, V. G. J. Stoks, and R. Schiavilla, 
  Phys. Rev. C {\bf 51}, 38 (1995).
\bibitem{N3LO} 
  E.~Epelbaum, H.~-W.~Hammer and U.~-G.~Mei{\ss}ner,
  Rev.\ Mod.\ Phys.\  {\bf 81}, 1773 (2009).

\bibitem{CioStr91} 
C. Ciofi degli Atti, S. Simula, L. L. Frankfurt, and M. I. Strikman,
Phys. Rev. C {\bf 44}, R7(R) (1991)
\bibitem{CioMezMor17a} 
C. Ciofi degli Atti, C. B. Mezzetti, and H. Morita
Phys. Rev. C {\bf 95}, 044327 (2017)
\bibitem{CioMor17b} 
C. Ciofi degli Atti and H. Morita
Phys. Rev. C {\bf 96}, 064317 (2017)

\bibitem{Arrington2011xs}
J. Arrington, D.W. Higinbotham, G. Rosner, and M. Sargsian,
Prog. Part. Nucl. Phys. {\bf 67}, 898 (2012)

\bibitem{ColleHen15}
C. Colle  {\it et al.},
Phys. Rev. C {\bf 92}, 024604 (2015)

\bibitem{Colle2015ena}
C. Colle, W. Cosyn, and J. Ryckebusch,
Phys. Rev. C {\bf 93}, 034608 (2016)

\bibitem{Boeglin2011mt}
W.U. Boeglin {\it et al.}, 
Phys. Rev. Lett. {\bf 107}, 262501 (2011)

\bibitem{Frankfurt1996xx}
L. L. Frankfurt, M. M. Sargsian, and M. I. Strikman,
Phys. Rev. C {\bf 56}, 1124 (1997)


\bibitem{Dutta2013}
D. Dutta, K. Hafidi, and M. Strikman,
Prog. Part. Nucl. Phys. {\bf 69}, 1 (2013)

\bibitem{Hen:2012yva}
O. Hen {\it et al.}, (CLAS Collaboration),
Phys. Lett. B {\bf 722}, 63 (2013)

\bibitem{Frankfurt2001}
L. Frankfurt, M. Strikman, and M. Zhalov,
Phys. Lett. B {\bf 503}, 73 (2001)

\bibitem{Pieper1992}
V. R. Pandharipande and S. C. Pieper,
Phys. Rev. C {\bf 45}, 791 (1992)

\bibitem{Ryc15} 
J. Ryckebusch {\it et al.}, 
J. Phys. G: Nucl. Part. Phys. {\bf 42}, 055104 (2015)
\bibitem{NeffFeldHor15} 
T. Neff, H. Feldmeier, and W. Horiuchi
Phys. Rev. C {\bf 92}, 024003 (2015)

\bibitem{Wiringa_CVMC} 
 D. Lonardoni, A. Lovato, S. C. Pieper, and R. B. Wiringa,
Phys. Rev. C {\bf 96}, 024326 (2017) 

\bibitem{ubx} 
  S. C. Pieper, V. R. Pandharipande, R. B. Wiringa, and J. Carlson,
  Phys. Rev. C {\bf 64}, 014001 (2001).

\bibitem{GezSch13} 
A. Gezerlis, I. Tews, E. Epelbaum, S. Gandolfi, K. Hebeler, A. Nogga, and A. Schwenk,
Phys. Rev. Lett. {\bf 111}, 032501 (2013)
\bibitem{GezSch14} 
A. Gezerlis, I. Tews, E. Epelbaum, M. Freunek, S. Gandolfi, K. Hebeler, A. Nogga, and A. Schwenk,
Phys. Rev. C {\bf 90}, 054323 (2014)

\bibitem{charge_density} 
R. Weiss, A. Schmidt, G. A. Miller, and N. Barnea,
Phys. Lett. B {\bf 790}, 484 (2019)

\bibitem{Chen17_EMC} 
J.-W. Chen, W. Detmold, J. E. Lynn, and A. Schwenk,
Phys. Rev. Lett. {\bf 119}, 262502 (2017)










%

%
%
%
%

%



\end{thebibliography}
\end{document}